\documentclass{article}

\usepackage{amssymb} %

\begin{document}
\title{Generation of solutions of the Hamilton--Jacobi equation}

\author{G.F.\ Torres del Castillo \\
Departamento de F\'isica Matem\'atica, Instituto de Ciencias \\
Universidad Aut\'onoma de Puebla, 72570 Puebla, Pue., M\'exico}

\maketitle

\begin{abstract}
It is shown that any function $G(q_{i}, p_{i}, t)$, defined on the extended phase space, defines a one-parameter group of canonical transformations which act on any function $f(q_{i}, t)$, in such a way that if $G$ is a constant of motion then from a solution of the Hamilton--Jacobi (HJ) equation one obtains a one-parameter family of solutions of the same HJ equation. It is also shown that any complete solution of the HJ equation can be obtained in this manner by means of the transformations generated by $n$ constants of motion in involution.
\end{abstract}

\noindent Keywords: Hamilton--Jacobi equation; canonical transformations; constants of motion 

\noindent PACS: 45.20.Jj; 02.30.Jr; 02.20.Qs

\section{Introduction}
In the Hamiltonian formulation of classical mechanics, the canonical transformations, given locally by expressions of the form $Q_{i} = Q_{i}(q_{j}, p_{j}, t)$, $P_{i} = P_{i}(q_{j}, p_{j}, t)$, act on the extended phase space (the Cartesian product of the phase space and the time axis) and, therefore, there is a naturally defined action of a canonical transformation on any function defined on the extended phase space (that is, on any function of $q_{i}$, $p_{i}$, and $t$). On the other hand, Hamilton's principal function, usually denoted by an $S$, is a function defined on the extended configuration space (that is, a function of $q_{i}$ and $t$), just like the wave function in the elementary Schr\"odinger equation, but there is no obvious way to define the action of a canonical transformation on a function of $q_{i}$ and $t$.

However, in Ref.\ \cite{DI} it was shown that, under certain conditions, one can define an action of a time-independent canonical transformation [that is, a canonical transformation that does not involve the time, $Q_{i} = Q_{i}(q_{j}, p_{j})$, $P_{i} = P_{i}(q_{j}, p_{j})$] on functions defined on the configuration space, in such a way that if a (time-independent) Hamiltonian is invariant under the canonical transformation, a solution of the corresponding time-independent Hamilton--Jacobi (HJ) equation is mapped into another solution. Furthermore, it was shown that the action of the one-parameter group of canonical transformations generated by an arbitrary function $G(q_{i}, p_{i})$ is determined by an equation similar to the HJ equation.

In this paper, we consider systems with a Hamiltonian that can depend on the time, we show that one can define the action of the one-parameter group of canonical transformations generated by an arbitrary function $G(q_{i}, p_{i}, t)$ on functions defined on the extended configuration space, in such a way that if $G$ is a constant of motion, then a solution of the corresponding HJ equation is mapped into a one-parameter family of solutions of this equation. In this manner, each constant of motion adds a continuous parameter to a given solution of the HJ equation. We also show that any complete solution of the HJ equation can be obtained from a solution without parameters by means of the action of the groups of canonical transformations generated by $n$ constants of motion in involution.

Throughout this paper, several examples are given in order to illustrate the definitions and results presented here.

\section{The action of a one-parameter group of canonical transformations on functions defined on the extended configuration space}
We start by reviewing the HJ equation, which will give us the pattern to follow in the definition of the action of any one-parameter group of canonical transformations on functions defined on the extended configuration space.

\subsection{The Hamilton--Jacobi as an evolution equation}
Given a Hamiltonian $H(q_{i}, p_{i}, t)$ of a system with $n$ degrees of freedom, the corresponding HJ equation is the first-order partial differential equation
\begin{equation}
H(q_{i}, \partial S/\partial q_{i}, t) + \frac{\partial S}{\partial t} = 0. \label{hj}
\end{equation}
Usually one is interested in {\em complete solutions}\/ of the HJ equation (\ref{hj}), that is, solutions $S(q_{i}, t, \alpha_{i})$ of Eq.\ (\ref{hj}), containing $n$ arbitrary parameters $\alpha_{i}$, such that
\begin{equation}
\det \left( \frac{\partial^{2} S}{\partial q_{i} \partial \alpha_{j}} \right) \not= 0, \label{compl}
\end{equation}
because they yield the solution of the Hamilton equations (see, {\em e.g.}, Refs.\ \cite{Wi} and \cite{Ga}). However, one can consider the HJ equation as an {\em evolution equation}, which determines the function $S(q_{i}, t)$ that reduces to a given function $f(q_{i})$, for $t = 0$ (or any other initial value, $t_{0}$, of $t$). (This fact does not sound very strange if we consider the relationship between the HJ equation and the Schr\"odinger equation.) In fact, if we have the solution of the Hamilton equations, we can use it to find the solution of the HJ equation that satisfies any initial condition $S(q_{i}, t_{0}) = f(q_{i})$. (Similar ideas are often mentioned in the standard textbooks on analytical mechanics, without presenting, however, a clear statement or an explicit procedure.)

For example, in the case of the Hamiltonian
\begin{equation}
H = \frac{p_{1}{}^{2} + p_{2}{}^{2}}{2m} + mg q_{2}, \label{hamex}
\end{equation}
where $m$ and $g$ are constants, the solution of the corresponding Hamilton equations can be readily obtained and is given by
\begin{equation}
q_{1} = (q_{1})_{0} + \frac{(p_{1})_{0} t}{m}, \qquad q_{2} = (q_{2})_{0} + \frac{(p_{2})_{0} t}{m} - \frac{gt^{2}}{2}, \qquad p_{1} = (p_{1})_{0}, \qquad p_{2} = (p_{2})_{0} - mgt, \label{solex}
\end{equation}
where $(q_{1})_{0}$ denotes the value of $q_{1}$ at $t = 0$, and so on. As the initial condition we choose
\begin{equation}
S(q_{1}, q_{2}, 0) = \alpha_{1} q_{1} + \alpha_{2} q_{2}, \label{icex}
\end{equation}
where $\alpha_{1}$, $\alpha_{2}$ are arbitrary constants. We want to find the solution of the HJ equation [see Eqs.\ (\ref{hj}) and (\ref{hamex})]
\begin{equation}
\frac{\partial S}{\partial t} = - \frac{1}{2m} \left[ \left( \frac{\partial S}{\partial q_{1}} \right)^{2} + \left( \frac{\partial S}{\partial q_{2}} \right)^{2} \right] - mg q_{2}, \label{hjex}
\end{equation}
that satisfies the initial condition (\ref{icex}). Recalling that $\partial S/\partial q_{i} = p_{i}$, making use of Eqs.\ (\ref{solex}) and (\ref{icex}) we have
\[
\frac{\partial S}{\partial q_{1}} = \left. \frac{\partial S}{\partial q_{1}} \right|_{t = 0} = \alpha_{1}, \qquad \frac{\partial S}{\partial q_{2}} = \left. \frac{\partial S}{\partial q_{2}} \right|_{t = 0} - mgt = \alpha_{2} - mgt.
\]
Substituting these expressions into the right-hand side of Eq.\ (\ref{hjex}) we have
\[
\frac{\partial S}{\partial t} = - \frac{1}{2m} \big[ \alpha_{1}{}^{2} + (\alpha_{2} - mgt)^{2} \big] - mg q_{2}.
\]
Combining the last three equations one readily finds that, choosing the integration constant so that Eq.\ (\ref{icex}) is satisfied,
\begin{equation}
S(q_{1}, q_{2}, t) = \alpha_{1} q_{1} + \alpha_{2} q_{2} - mgtq_{2} - \frac{\alpha_{1}{}^{2} t}{2m} + \frac{(\alpha_{2} - mgt)^{3} - \alpha_{2}{}^{3}}{6m^{2}g}. \label{prinex}
\end{equation}
The expression (\ref{prinex}) is a (complete, R-separable) solution of the HJ equation that reduces to the specified function (\ref{icex}) for $t = 0$. (A solution is R-separable if it is the sum of a function of two or more variables and functions of one variable.)

We close this subsection with a second example. The solution of the Hamilton equations for the time-dependent Hamiltonian
\begin{equation}
H = \frac{p^{2}}{2m} - ktq, \label{n1}
\end{equation}
where $m$ and $k$ are constants, is given by
\begin{equation}
q = q_{0} + \frac{tp_{0}}{m} + \frac{kt^{3}}{6m}, \qquad p = p_ {0} + \frac{1}{2} kt^{2}, \label{n2}
\end{equation}
where $q_{0}$ and $p_{0}$ denote the values of $q$ and $p$ at $t = 0$, respectively. Letting
\begin{equation}
S(q,0,\alpha) = \alpha q, \label{n3}
\end{equation}
where $\alpha$ is an arbitrary constant, with the aid of Eqs.\ (\ref{n2}), we have
\begin{equation}
\frac{\partial S}{\partial q} = \left. \frac{\partial S}{\partial q} \right|_{t = 0} + \frac{1}{2} kt^{2} = \alpha + \frac{1}{2} kt^{2}. \label{n4}
\end{equation}
Thus, from Eqs.\ (\ref{hj}), (\ref{n1}), and (\ref{n4}), we have
\begin{equation}
\frac{\partial S}{\partial t} = - \frac{1}{2m} \left( \frac{\partial S}{\partial q} \right)^{2} + ktq = - \frac{1}{2m} \left( \alpha + \frac{1}{2} kt^{2} \right)^{2} + ktq. \label{n5}
\end{equation}
Then, from Eqs.\ (\ref{n3})--(\ref{n5}), we readily obtain
\[
S(q, t, \alpha) = \alpha q + \frac{1}{2} kt^{2} q - \frac{1}{2m} \left( \alpha^{2} t + \frac{1}{3} \alpha k t^{3} + \frac{k^{2} t^{5}}{20} \right),
\]
which is an R-separable complete solution of the HJ equation. (We might start with expressions more complicated than (\ref{icex}) and (\ref{n3}), but these simple expressions are enough to obtain complete solutions of the HJ equation.)

It should be clear that a similar construction can be devised using an arbitrary function of $(q_{i}, p_{i}, t)$ instead of a Hamiltonian.

\subsection{Families of solutions of the HJ equation and constants of motion}
Any differentiable function, $G(q_{i}, p_{i}, t)$, defines a (possibly local) one-parameter group of canonical transformations, determined by the (autonomous) system of first-order ordinary differential equations
\begin{equation}
\frac{{\rm d} q_{i}}{{\rm d} \alpha} = \frac{\partial G(q_{j}, p_{j}, t)}{\partial p_{i}}, \qquad
\frac{{\rm d} p_{i}}{{\rm d} \alpha} = - \frac{\partial G(q_{j}, p_{j}, t)}{\partial q_{i}}, \label{odes}
\end{equation}
which are of the form of the Hamilton equations.

For example, the function
\begin{equation}
G(q_{1}, q_{2}, p_{1}, p_{2}, t) = \frac{p_{1} p_{2}}{m} + mgq_{1}, \label{a}
\end{equation}
where $m$ and $g$ are constants, leads to the system of equations
\begin{equation}
\frac{{\rm d} q_{1}}{{\rm d} \alpha} = \frac{p_{2}}{m}, \qquad \frac{{\rm d} q_{2}}{{\rm d} \alpha} = \frac{p_{1}}{m}, \qquad \frac{{\rm d} p_{1}}{{\rm d} \alpha} = - mg, \qquad \frac{{\rm d} p_{2}}{{\rm d} \alpha} = 0. \label{sim}
\end{equation}
From the last two equations we find that
\begin{equation}
p_{1} = (p_{1})_{0} - mg \alpha, \qquad p_{2} = (p_{2})_{0}, \label{apes}
\end{equation}
where $(p_{i})_{0}$ denotes the value of $p_{i}$ for $\alpha = 0$. Substituting these expressions into the first pair of equations (\ref{sim}) we get
\[
\frac{{\rm d} q_{1}}{{\rm d} \alpha} = \frac{(p_{2})_{0}}{m}, \qquad \frac{{\rm d} q_{2}}{{\rm d} \alpha} = \frac{(p_{1})_{0} - mg \alpha}{m}.
\]
Hence,
\begin{equation}
q_{1} = (q_{1})_{0} + \frac{(p_{2})_{0} \alpha}{m}, \qquad q_{2} = (q_{2})_{0} + \frac{(p_{1})_{0} \alpha}{m} - \frac{g \alpha^{2}}{2}. \label{aqs}
\end{equation}

The action of the canonical transformations defined by a function $G(q_{i}, p_{i}, t)$, on functions of
$(q_{i}, t)$, will be defined imitating the HJ equation.

\noindent{\bf Definition.} The image of a given arbitrary function $f(q_{i}, t)$ under the one-parameter group of canonical transformations generated by $G(q_{i}, p_{i}, t)$ is the function $S(q_{i}, t, \alpha)$ such that
\begin{equation}
G(q_{i}, \partial S/\partial q_{i}, t) + \frac{\partial S}{\partial \alpha} = 0, \label{action}
\end{equation}
with the initial condition $S(q_{i}, t, 0) = f(q_{i}, t)$. ({\em Cf.}\ Eq.\ (\ref{hj}).)

The following example shows that this definition produces the expected effect in the case of transformations that only affect the coordinates of the configuration space. Indeed, the one-parameter group of canonical transformations generated by, {\em e.g.}, $G = p_{1}$ is given by [see Eqs.\ (\ref{odes})]
\[
q_{1} = (q_{1})_{0} + \alpha, \qquad q_{i} = (q_{i})_{0}, \quad {\rm for \ } i \geqslant 2, \qquad p_{i} = (p_{i})_{0}, \]
that is, ``translations'' along the $q_{1}$-axis. Then, Eq.\ (\ref{action}) takes the form
\[
\frac{\partial S}{\partial q_{1}} + \frac{\partial S}{\partial \alpha} = 0,
\]
whose general solution is
\[
S(q_{i}, t, \alpha) = F(q_{1} - \alpha, q_{2}, \ldots, q_{n}, t),
\]
where $F$ is an arbitrary function. By imposing the initial condition $S(q_{i}, t, 0) = f(q_{i}, t)$, we find that
\[
S(q_{i}, t, \alpha) = f(q_{1} - \alpha, q_{2}, \ldots, q_{n}, t),
\]
which is the expected effect of a translation along the $q_{1}$-axis.

As a second example, the images of the function
\begin{equation}
f(q_{i}, t) = (\alpha_{2} - mgt)q_{2} + \frac{(\alpha_{2} - mgt)^{3} - \alpha_{2}{}^{3}}{6m^{2}g} \label{icex2}
\end{equation}
(which is the function (\ref{prinex}) with $\alpha_{1} = 0$) under the transformations generated by (\ref{a}) can be obtained proceeding as in the examples of Sec.\ 2.1. Making use of Eqs.\ (\ref{apes}) and (\ref{icex2}) we have [with $S(q_{i}, t, 0) = f(q_{i}, t)$]
\begin{equation}
\frac{\partial S}{\partial q_{1}} = \left. \frac{\partial S}{\partial q_{1}} \right|_{\alpha = 0} - mg \alpha = - mg \alpha, \qquad \frac{\partial S}{\partial q_{2}} = \left. \frac{\partial S}{\partial q_{2}} \right|_{\alpha = 0} = \alpha_{2} - mgt. \label{partial}
\end{equation}
Substituting these expressions into Eq.\ (\ref{action}), taking into account Eq.\ (\ref{a}), we obtain
\[
\frac{\partial S}{\partial \alpha} = - \frac{1}{m} (- mg \alpha) (\alpha_{2} - mgt) - mg q_{1}.
\]
Combining the last equations and the initial condition (\ref{icex2}) one finds that
\begin{equation}
S(q_{1}, q_{2}, t, \alpha_{2}, \alpha) = (\alpha_{2} - mgt)q_{2} - mg \alpha q_{1} + \frac{1}{2} (\alpha_{2} - mgt)g \alpha^{2} + \frac{(\alpha_{2} - mgt)^{3} - \alpha_{2}{}^{3}}{6m^{2}g}. \label{prinex2}
\end{equation}
Note that $t$ is treated here as a constant because the canonical transformations do not affect $t$. Since the function (\ref{icex2}) is a particular case of (\ref{prinex}), is a solution of the HJ equation (\ref{hjex}), and one can readily verify that (\ref{prinex2}) is {\em also}\/ a solution of Eq.\ (\ref{hjex}), for all values of the parameter $\alpha$ [which is essentially the two-parameter family of solutions (\ref{prinex})].

Thus, we have obtained a two-parameter family of solutions of the HJ equation (\ref{hjex}), starting from a one-parameter solution of this equation. As we shall show in the following Proposition, this is a consequence of the fact that the function (\ref{a}) is a constant of motion for the Hamiltonian (\ref{hamex}).

\noindent {\bf Proposition 1.} The image of a solution of the HJ equation, $S_{0}(q_{i}, t)$, under the one-parameter group of canonical transformations generated by a constant of motion $G(q_{i}, p_{i}, t)$ is a one-parameter family of solutions of the same HJ equation.

\noindent {\em Proof.} Assuming that $S(q_{i}, t, \alpha)$ is a solution of Eq.\ (\ref{action}), making use of the chain rule repeatedly, we have (here, and in what follows, there is summation over repeated indices)
\begin{eqnarray}
\frac{\partial}{\partial \alpha} \left[ H(q_{i}, \partial S/\partial q_{i}, t) + \frac{\partial S}{\partial t} \right] & = & \frac{\partial H}{\partial p_{j}} \frac{\partial^{2} S}{\partial \alpha \partial q_{j}} + \frac{\partial^{2} S}{\partial \alpha \partial t} \nonumber \\
& = & - \frac{\partial H}{\partial p_{j}} \frac{\partial G(q_{i}, \partial S/\partial q_{i}, t)}{\partial q_{j}} - \frac{\partial G(q_{i}, \partial S/\partial q_{i}, t)}{\partial t} \nonumber \\
& = & - \frac{\partial H}{\partial p_{j}} \left( \frac{\partial G}{\partial q_{j}} + \frac{\partial G}{\partial p_{i}} \frac{\partial^{2} S}{\partial q_{j} \partial q_{i}} \right) - \frac{\partial G}{\partial p_{i}} \frac{\partial^{2} S}{\partial t \partial q_{i}} - \frac{\partial G}{\partial t}. \label{des1}
\end{eqnarray}
According to the hypothesis, $G$ is a constant of motion, that is
\begin{equation}
\frac{\partial G}{\partial t} + \frac{\partial G}{\partial q_{j}} \frac{\partial H}{\partial p_{j}} - \frac{\partial G}{\partial p_{j}} \frac{\partial H}{\partial q_{j}} = 0, \label{const}
\end{equation}
thus, the last line of Eq.\ (\ref{des1}) amounts to
\begin{eqnarray*}
- \frac{\partial G}{\partial p_{i}} \left( \frac{\partial H}{\partial q_{i}} + \frac{\partial H}{\partial p_{j}} \frac{\partial^{2} S}{\partial q_{j} \partial q_{i}} + \frac{\partial^{2} S}{\partial t \partial q_{i}} \right) & = & - \frac{\partial G}{\partial p_{i}} \frac{\partial}{\partial q_{i}} \left[ H(q_{i}, \partial S/\partial q_{i}, t) + \frac{\partial S}{\partial t} \right] \\
& = & - \frac{{\rm d} q_{i}}{{\rm d} \alpha} \frac{\partial}{\partial q_{i}} \left[ H(q_{i}, \partial S/\partial q_{i}, t) + \frac{\partial S}{\partial t} \right]
\end{eqnarray*}
[see Eqs.\ (\ref{odes})]. Hence, we obtain the equation
\[
\frac{{\rm d}}{{\rm d} \alpha} \left[ H(q_{i}(\alpha), \partial S(q_{i}(\alpha), t, \alpha)/\partial q_{i}, t) + \frac{\partial S(q_{i}(\alpha), t, \alpha)}{\partial t} \right] = 0,
\]
which involves the images of $S$ and the coordinates $q_{i}$ under the transformations generated by $G$. Since the expression inside the brackets in the last equation vanishes for $\alpha = 0$, it is equal to zero for all values of $\alpha$, thus proving the assertion.

A partial converse of this Proposition is true. The steps in the proof show that if the images of a solution of the HJ equation under the transformations generated by $G$ are solutions of the same HJ equation, and $S$ is a complete solution of the HJ equation, then $G$ is a constant of motion. The assumption that $S$ is a complete solution of the HJ equation assures that Eq.\ (\ref{const}) holds everywhere. In fact, the following stronger result holds.

\noindent {\bf Proposition 2.} A complete solution of the HJ equation, $S(q_{i}, t, \alpha_{i})$ (not necessarily separable or R-separable), is the image of $S(q_{i}, t, 0)$ under the transformations generated by the $n$ constants of motion
\begin{equation}
G_{j}(q_{i}, p_{i}, t) = - \frac{\partial S}{\partial \alpha_{j}}(q_{i}, t, \alpha_{i}(q_{k}, p_{k}, t)) \label{prop3}
\end{equation}
($j = 1, 2, \ldots, n$), obtained by eliminating the $\alpha_{i}$ by means of
\begin{equation}
p_{j} = \frac{\partial S}{\partial q_{j}} (q_{i}, \alpha_{i}, t). \label{pes}
\end{equation}
Furthermore, the functions $G_{i}$ are in involution ({\em i.e.}, the Poisson bracket of $G_{i}$ and $G_{j}$ is equal to zero). (Note that, according to Eq.\ (\ref{compl}), in principle, one can find the $\alpha_{i}$ from Eqs.\ (\ref{pes}).)

\noindent {\em Proof.} We note that, by virtue of Eqs.\ (\ref{pes}), Eq.\ (\ref{prop3}) is equivalent to Eq.\ (\ref{action}). Since, by hypothesis, $S(q_{i}, t, \alpha_{i})$ is a complete solution of the HJ equation, each function $G_{i}$, defined by (\ref{prop3}), is a constant of motion. In order to prove that the functions $G_{i}$ are in involution, making use of Eq.\ (\ref{prop3}), we calculate the mixed partial derivative
\[
\frac{\partial^{2} S}{\partial \alpha_{i} \partial \alpha_{j}} = - \frac{\partial}{\partial \alpha_{i}} G_{j}(q_{i}, \partial S/\partial q_{i}, t, \alpha_{k}) = - \frac{\partial G_{j}}{\partial p_{k}} \frac{\partial}{\partial \alpha_{i}} \frac{\partial S}{\partial q_{k}} = \frac{\partial G_{j}}{\partial p_{k}} \frac{\partial G_{i}}{\partial q_{k}}.
\]
Then we see that the commutativity of the partial derivatives of $S$ implies that the Poisson brackets of the $G_{i}$ are equal to zero. (The assumption that $S(q_{i}, t, \alpha_{i})$ is a complete solution of the HJ equation implies that the Poisson brackets $\{ G_{i}, G_{j} \}$ vanish everywhere.) Since the Poisson brackets $\{ G_{i}, G_{j} \}$ are equal to zero, the one-parameter groups of transformations generated by the functions $G_{i}$ commute. (In fact, the groups of canonical transformations generated by two functions $F$ and $G$ commute if and only if $\{ F, G \}$ is a trivial constant, not necessarily zero \cite{DM}.)

Let us consider, for example, the function
\begin{equation}
S(q_{i}, t, \alpha_{i}) = \alpha_{1} q_{1} + \alpha_{2} q_{2} - mgtq_{2} - \frac{\alpha_{1}{}^{2} t}{2m} + \frac{(\alpha_{2} - mgt)^{3} - \alpha_{2}{}^{3}}{6m^{2}g}, \label{prinexo}
\end{equation}
which is the complete solution (\ref{prinex}) of the HJ equation constructed in Sec.\ 2.1. One finds that
\[
\frac{\partial S}{\partial q_{1}} = \alpha_{1}, \qquad \frac{\partial S}{\partial q_{2}} = \alpha_{2} - mgt,
\]
which lead to [see Eqs.\ (\ref{pes})]
\begin{equation}
\alpha_{1} = p_{1}, \qquad \alpha_{2} = p_{2} + mgt. \label{inversi}
\end{equation}
On the other hand,
\[
\frac{\partial S}{\partial \alpha_{1}} = q_{1} - \frac{\alpha_{1} t}{m}, \qquad \frac{\partial S}{\partial \alpha_{2}} = q_{2} + \frac{(\alpha_{2} - mgt)^{2} - \alpha_{2}{}^{2}}{2 m^{2} g},
\]
which, taking into account Eqs.\ (\ref{inversi}), are of the form (\ref{prop3}) with
\begin{equation}
G_{1} = - q_{1} + \frac{p_{1} t}{m}, \qquad G_{2} = - q_{2} + \frac{p_{2} t}{m} + \frac{1}{2} gt^{2}. \label{ges}
\end{equation}
One can readily verify that $G_{1}$ and $G_{2}$ are constants of motion in involution, and that the function (\ref{prinexo}) can be obtained by means of the transformations generated by $G_{1}$ and $G_{2}$ from
\begin{equation}
S(q_{i},t) = - mgt q_{2} - \frac{mg^{2}t^{3}}{6}, \label{init}
\end{equation}
which is obtained from (\ref{prinexo}) setting $\alpha_{1} = \alpha_{2} = 0$.

Indeed, the function
\begin{equation}
G \equiv \alpha_{1} G_{1} + \alpha_{2} G_{2} = \alpha_{1} \left( - q_{1} + \frac{p_{1} t}{m} \right) + \alpha_{2} \left( - q_{2} + \frac{p_{2} t}{m} + \frac{1}{2} gt^{2} \right) \label{gee}
\end{equation}
is a constant of motion for all values of the constants $\alpha_{1}$ and $\alpha_{2}$. The one-parameter group of canonical transformations generated by $G$ is given by (denoting by $s$ the corresponding parameter)
\[
q_{1} = (q_{1})_{0} + \frac{\alpha_{1} t}{m} s, \quad q_{2} = (q_{2})_{0} + \frac{\alpha_{2} t}{m} s, \quad p_{1} = (p_{1})_{0} + \alpha_{1} s, \quad p_{2} = (p_{2})_{0} + \alpha_{2} s,
\]
hence, making use of the initial condition (\ref{init}),
\[
\frac{\partial S}{\partial q_{1}} = 0 + \alpha_{1} s, \qquad \frac{\partial S}{\partial q_{2}} = - mgt + \alpha_{2} s,
\]
and substituting these expressions into Eq.\ (\ref{action}), with $G$ given by (\ref{gee}),
\[
\frac{\partial S}{\partial s} = \alpha_{1} q_{1} - \frac{\alpha_{1}t}{m}(\alpha_{1} s) + \alpha_{2} q_{2} - \frac{\alpha_{2} t}{m} (-mgt + \alpha_{2} s) - \frac{1}{2} \alpha_{2} gt^{2}.
\]
Thus,
\[
S(q_{1}, q_{2}, t) = \alpha_{1} s q_{1} + \alpha_{2} s q_{2} - mgtq_{2} - \frac{(\alpha_{1} s)^{2} t}{2m} + \frac{(\alpha_{2} s - mgt)^{3} - (\alpha_{2} s)^{3}}{6m^{2}g},
\]
which reduces to the expression (\ref{prinexo}) when $s = 1$.

\section{Concluding remarks}
According to Proposition 2, any complete solution of the HJ equation is, locally, the image of a particular solution (without free parameters) under the action of an Abelian $n$-dimensional group (that is, a group locally isomorphic to the additive Lie group $\mathbb{R}^{n}$). As pointed out above, these solutions need not be separable or R-separable. It should be noted, however, that some functions are invariant under the transformations generated by a function $G$, even if these transformations are different from the identity. For instance, a function that does not depend on $q_{1}$ is invariant under the translations generated by $p_{1}$.

One can readily see that, at least in the case of time-independent operators, an analog of Proposition 1 holds in the case of the solutions of the Schr\"odinger equation.

In addition to the R-separable complete solution (\ref{prinex}), the HJ equation for the Hamiltonian (\ref{hamex}) also admits complete separable solutions in the same coordinate system (see also Ref.\ \cite{Li})


\begin{thebibliography}{9}
\bibitem{DI} G.F.\ Torres del Castillo, D.A.\ Rosete \'Alvarez and I.\ Fuentecilla C\'arcamo, {\it Rev.\ Mex.\ F\'is.}\ {\bf 56} (2010) 113.
\bibitem{Wi} E.T.\ Whittaker, A Treatise on the Analytical Dynamics of Particles and Rigid Bodies, 4th ed.\ (Cambridge University Press, Cambridge, 1993). Chap.\ XII.
\bibitem{Ga} F.\ Gantmacher, Lectures in Analytical Mechanics (Mir, Moscow, 1975). Chap.\ 4.
\bibitem{DM} G.F.\ Torres del Castillo, {\it Differentiable Manifolds: A Theoretical Physics Approach} (Birkh\"auser Science, New York, 2012). Sec.\ 8.2.
\bibitem{Li} G.F.\ Torres del Castillo, {\it Rev.\ Mex.\ F\'is.}\ {\bf 59} (2013) 478.
\end{thebibliography}
\end{document}